\documentstyle[prl, aps, amssymb]{revtex}
\begin{document}
\draft
\wideabs{
\title{Systematic $^{63}$Cu NQR studies of the stripe phase in
La$_{1.6-x}$Nd$_{0.4}$Sr$_{x}$CuO$_{4}$ for 0.07$\leq x \leq $0.25}
\author{P.M. Singer, A.W. Hunt, A.F. Cederstr\"om, and T. Imai}
\address{Department of Physics and Center for Materials Science and 
Engineering, Massachusetts Institute of Technology, Cambridge, MA 02139}
\date{\today}
\maketitle
\begin{abstract}
We demonstrate that the integrated intensity of $^{63}$Cu nuclear 
quadrupole resonance (NQR) in La$_{1.6-x}$Nd$_{0.4}$Sr$_{x}$CuO$_{4}$
decreases dramatically below the charge-stripe ordering temperature 
$T_{charge}$. Comparison with neutron and X-ray scattering indicates 
that the wipeout fraction $F(T)$ (i.e. the missing fraction of the 
integrated intensity of the NQR signal) represents the charge-stripe 
order parameter. The systematic study reveals bulk charge-stripe 
order throughout the superconducting region $0.07\leq x\leq 0.25$.
As a function of the reduced temperature $t\equiv T/T_{charge}$,  
the temperature dependence of $F(t)$ is sharpest for the hole 
concentration $x\sim1/8$, indicating that $x\sim1/8$ is the optimum 
concentration for stripe formation.  

\end{abstract}
}

\section{introduction}

Research into the stripe phase of the CuO$_{2}$ plane 
has continued to expand \cite{hunt,suzukip,kimura,waki,lee} since its experimental discovery 
in 1995 by Tranquada {\it et al.} \cite{tranquada1}.
Static charge-stripe order was observed for
La$_{1.6-x}$Nd$_{0.4}$Sr$_{x}$CuO$_{4}$ with $x=0.12$ \cite{tranquada2} 
with an onset temperature $T_{charge}\sim 65$K
using neutron diffraction. The same charge-order superlattice peaks  
were recently confirmed 
 using hard X-ray 
 diffraction by Zimmerman {\it et al.} \cite{zimmerman} and a similar 
$T_{charge}\sim 70$K was found.  
 Even more recently, charge-order superlattice peaks have
  been observed in La$_{1.45}$Nd$_{0.4}$Sr$_{0.15}$CuO$_{4}$ by 
 Niem\"oller {\it et al.} \cite{niemoller} using hard X-rays,
 and they report a slightly lower onset temperature $T_{charge}\sim 62$K. 
Charge transport studies of 
La$_{1.6-x}$Nd$_{0.4}$Sr$_{x}$CuO$_{4}$ by Uchida {\it et al.} \cite{uchida}
also support a 
static charge-stripe picture. The Hall coefficient 
shows a dramatic decrease starting around $T_{charge}$,
reflecting the one dimensional nature of the charge transport in 
the striped phase. 
Furthermore, the decrease is sharpest around the $x=1/8$ doping where 
stripe order is believed to be most robust \cite{tranquada1}. 

The observation of spin-stripe order 
 has been reported for an even wider 
range of doping. Neutron scattering by Tranquada {\it et al.} 
\cite{tranquada3} successfully detected static 
spin-stripe order for $x=0.12,0.15$ and $0.20$ at $T_{spin}\sim$ 50K, 
45K, and 20K respectively. These findings have been comfirmed and 
extended with more recent measurements by Ichikawa {\it et al.} \cite{ichikawa}.
The neutron scattering experiments \cite{tranquada3} also observe a decreasing 
sublattice magnetization away from $1/8$ doping, and
incommensurabilities that are similar to those for the
 La$_{2-x}$Sr$_{x}$CuO$_{4}$ 
\cite{yamada} series. The most interesting feature however, is that the 
onset temperature $T_{spin}$ for spin-stripe order 
in La$_{1.6-x}$Nd$_{0.4}$Sr$_{x}$CuO$_{4}$ at $x=0.12$ is 50K 
, which is $lower$ than $T_{charge}$ of 65K, implying that 
charge order is a precursor to spin order. 
In addition, a spin ordering temperature $T_{spin}=$ 30K, 25K, 
and $\leq$ 4K for $x = 0.12$, 0.15, and 0.20 respectively has been determined by 
Nachumi {\it et al.} \cite{nachumi} using $\mu$SR. 
The $\mu$SR probe has a lower inherent frequency 
($\sim 10^{7}$ Hz) compared with elastic neutron scattering ($\sim 
10^{11}$ Hz), so the discrepency in the spin ordering temperature 
between the $\mu$SR and neutron results indicate that the spin-stripe 
fluctuations
 gradually slow down with decreasing temperature below $T_{charge}$.
 That is, spin-stripe ordering is a {\it glassy} transition in La$_{1.6-x}$Nd$_{0.4}$Sr$_{x}$CuO$_{4}$.
 
In this paper, we utilize Cu nuclear quadrupole resonance (NQR) to probe 
stripe instabilities in La$_{1.6-x}$Nd$_{0.4}$Sr$_{x}$CuO$_{4}$. Cu NQR 
gives unique information about the charge environment at the Cu nuclear 
site. In particular, the resonance frequency for Cu NQR, conventionally 
called $\nu_{Q}$ ($\sim 36$ MHz) \cite{abragam,slichter}, is 
determined by the energy splitting of the $\pm \frac {3} {2}$ 
$\leftrightarrow$ $\pm 
\frac {1} {2}$ levels, which in turn is directly proportional to 
the electric field gradient (EFG) at the Cu nuclear site. The EFG 
itself is very sensitive to the charge environment, and we thus expect 
any change in the charge distribution within the CuO$_{2}$ plane, 
such as charge density waves (CDW), to directly affect Cu NQR. 
Earlier work on conventional CDW systems such as NbSe$_{3}$ 
\cite{ross} and 
Rb$_{0.3}$MoO$_{3}$ \cite{nomura} made use of the extreme
sensitivity of $\nu_{Q}$ to the 
EFG. Our novel approach \cite{hunt} makes use of wipeout 
effects \cite{winter} in $^{63}$Cu NQR from which we obtain unique 
information about charge-stripe order in La$_{1.6-x}$Nd$_{0.4}$Sr$_{x}$CuO$_{4}$. 
The fundamental difference between NQR and neutron scattering is 
that the former is a local probe, while the latter requires spatial 
coherence, thus NQR and neutron scattering provide  
 complimentary information about stripe 
 physics. 
The NQR wipeout alone cannot give 
details of the spatial structure of the stripes, but being a local 
probe, NQR is very sensitive to charge-stripes with  
short or poorly defined correlation lengths. 
Indeed, away from the 
robust $1/8$ region, detection of charge-stripes by bulk scattering
 has proved difficult and
  up to now, no direct observation of 
charge-stripe order has been reported other than for
La$_{1.48}$Nd$_{0.4}$Sr$_{0.12}$CuO$_{4}$ and
La$_{1.45}$Nd$_{0.4}$Sr$_{0.15}$CuO$_{4}$.
However by confirming that the NQR wipeout fraction 
 $F(T)$ has {\it identical} temperature dependence 
 as the neutron and X-ray charge-stripe order parameter for $x=0.12$ 
 and 0.15, we claim that $F(T)$ is the charge-stripe order parameter and
we extend the detection of charge-stripe 
order in La$_{1.6-x}$Nd$_{0.4}$Sr$_{x}$CuO$_{4}$
to $0.07\leq x\leq 0.25$. We also establish that the transition is 
sharpest at $x=0.12$.  

The rest of the  paper is presented as follows: in section II we
 go over the experimental details and present 
 our 
results. In  
section III we discuss our NQR data in comparison 
with the neutron and X-ray measurements.
Section IV contains the conclusions.

\section{EXPERIMENTAL procedures and RESULTS}

\subsection{Sample preparation}

Powder samples of La$_{1.6-x}$Nd$_{0.4}$Sr$_{x}$CuO$_{4}$
 with $x=0.07,0.09,0.12,0.15,0.20,$ and $0.25$ were prepared by 
solid-state reactions of La$_{2}$O$_{3}$ $(99.99 \% )$,
Nd$_{2}$O$_{3}$ $(99.99 \% )$, SrCO$_{3}$ $(99.99 \% )$ and CuO $(99.995 \% )$.
The materials were mixed in the desired stochiometry and prereacted 
at $850^{\circ}$ C. 
The samples were then finely ground and heated to $950^{\circ}$ C. 
This cycle was repeated several times. 
For the final reaction, the powder was pressed (0.6 GPa) 
into rods of cylindrical 
shape and annealed in flowing oxygen. The heat cycle for this final
 reaction was similar to that used by Breuer {\it et al.} \cite{breuer}
 where 
a highest temperature of $1150^{\circ}$ C was used. All the samples 
were confirmed by X-ray diffraction to be single phased, and the room 
temperature orthorhombic splitting $[b-a]$ was in good agreement
with \cite{breuer}. 

\subsection {Characterization of NQR spectra}

The $^{63,65}$Cu NQR frequency spectra, or lineshapes, were measured 
using a $90^{\circ}-\tau -180^{\circ}-\tau- echo $ 
phase-cycled pulse-sequence at fixed $\tau$. The resonant circuit was 
heavily {\it damped} so that short $\tau$ could be used (typically $\tau = 
10 \sim 12 \mu s$). In addition, the damping assured that the {\it Q} 
value of the resonance circuit changed little with temperature, 
therefore pulse conditions and sensitivity remained constant.

Fig. \ref{1}(a) shows the lineshapes 
for all the La$_{1.6-x}$Nd$_{0.4}$Sr$_{x}$CuO$_{4}$ materials 
observed at 150K. At 150K,
all the materials are in the LTO (low temperature orthorhombic) phase. 
The two different sites, conventionally called A and B, are present 
for all the materials and each is further 
split into two lines due to the presence of $^{63}$Cu and $^{65}$Cu 
isotopes.
 Note that the intensity of the B 
line compared with that of the A line goes roughly as the doping 
$x$, in good agreement with earlier 
reports by Yoshimura {\it et al.} \cite{yoshimura}. 
It is generally believed that the B line 
originates from Cu nuclei underneath a Sr atom, and the A line 
from the Cu nuclei away from the Sr dopants.
 This picture is consistent with the 
observed 
 ratio of intensities. We also observed an enhanced tail at the lower frequency 
 side of the spectrum, possibly from Cu sites underneath a Nd atom. 
 This lower frequency tail is also observed in 
 La$_{1.8-x}$Eu$_{0.2}$Sr$_{x}$CuO$_{4}$ spectra \cite{hunt2}, 
 presumably from Cu sites underneath a Eu atom.

In order to extract quantitative information about the lineshapes, 
each spectrum was fit to a convolution of Gaussians; two  
for the $^{63}$Cu and $^{65}$Cu A line, and two for the $^{63}$Cu and 
$^{65}$Cu
B line (as shown for $x=0.12 $ in Fig. \ref{1}(a)).
Fitting the A line with
two Gaussians results in six parameters (two amplitudes, two peak 
positions corresponding to the NQR frequencies $^{63}\nu_{Q}$
 and $^{65}\nu_{Q}$, and 
two widths). Using the known ratios of 
$^{63}\nu_{Q}/^{65}\nu_{Q} = Q^{63}/Q^{65}=1.083$ 
and $N^{63}/N^{65}=69/31$ for the quadrupole moments $Q$ and isotopic 
abundances $N$ respectively reduces the number of free parameters. 
The B line was fit in the same manner.

Results of A line HWHM (half 
width at half maximum) observed at 150K are shown in Fig. \ref{1}(b).
The fitted value of the HWHM  
is sensitive to the lineshape data, especially at the tails,
but the general trend is that the HWHM increases with $x$. 
The value is generally $50 \%$  higher than the HWHM observed 
for La$_{2-x}$Sr$_{x}$CuO$_{4}$ at the same doping \cite{hunt}.
This is possibly due to the
increase in structural disorder due to Nd substitution.

Results of the
$^{63}\nu_{Q}$ for the A and B line observed at 150K are shown in Fig. 
\ref{1}(c). 
The doping dependence of $\nu_{Q}$ for the A and B line in
 La$_{1.6-x}$Nd$_{0.4}$Sr$_{x}$CuO$_{4}$ is consistent 
with previous studies on La$_{2-x}$Sr$_{x}$CuO$_{4}$ \cite{yoshimura}.

\subsection {Temperature dependence of $^{63}$Cu NQR spectra}

The temperature dependence of the lineshapes came in three forms: 

(1) The inhomogeneous linewidth (HWHM) of the $^{63}$Cu A line
shown in Fig. \ref{12_temp}(a) for $x=0.12$
shows a smooth increase ($\sim 
20 \%$) from 
300K to 20K. The same temperature dependence of the HWHM
 was found for all the 
samples. The HWHM data was used to estimate the integrated intensity of the
NQR 
lineshape, so to avoid unnecessary scatter in this estimation,
 we typically used a linear fit to the HWHM.

(2) Fig. \ref{12_temp}(a) shows the temperature dependence of 
$^{63}\nu_{Q}$ for the $x=0.12$ sample, and as can be seen,   
$^{63}\nu_{Q}$ shows a gradual increase of several hundred kHz
from room temperature 
down to $T_{charge}=$70(7)K, below which the rise in $^{63}\nu_{Q}$ is 
more dramatic. 
The other
samples show qualitatively similar temperature dependence
as discussed in section III.

(3) All of the materials showed a dramatic loss of signal intensity at 
temperatures below 150K. Fig. \ref{12_temp}(b) shows
the temperature dependence of the lineshapes for 
$x = 0.12$. The NQR intensity is 
proportional to the statistical Boltzmann factor $e^{h\nu_{Q} 
/k_{B}T} \sim 1/T$ 
where 
$h\nu_{Q} \ll k_{B}T$ in the present case. Accordingly, we multiplied each 
lineshape by $T$ to take this into account. The loss of NQR intensity 
therefore
implies that $^{63,65}$Cu nuclear spins in certain segments of the 
sample become undetectable.

In order to quantify the loss of signal shown in Fig.\ref{12_temp}(b),
 we estimated the temperature dependence of the integrated intensity
for each material 
using a three step process. (1) We calculated the area of the lineshape 
from the fitting procedure described above.  
(2) A Boltzmann factor was included at 
each temperature. (3) The lineshapes were corrected for spin echo decay 
as described below. 
We also attempted to make standard NQR frequency corrections 
of $1/ f^{2}$ to the lineshape intensity \cite{slichter}, but they 
did not affect our results. 
Fig. \ref{S}(a) shows the results of the 
temperature dependence of the integrated intensity
for $x=0.12$.
When working out the lineshape area, only the integral of the 
A line intensity was used so that unecessary scattering of the 
estimated total intensity was avoided.
This was justified by 
previous high precision studies of the NQR 
intensity \cite{hunt} on a related compound
La$_{1.875}$Ba$_{0.125}$CuO$_{4}$ enriched with the $^{63}$Cu isotope.
As shown in Fig. 
\ref{S}(b), both A and B lines have identical temperature dependence of the 
integrated NQR intensity.

Corrections for spin echo decay, or the `$T_{2}$ corrections',
were made by measuring the spin-spin 
relaxation $1/T_{2}$ \cite{slichter} at each temperature. 
Since $T_{2}$ exhibits slight dependence on frequency, in order to be rigorous 
one needs to make 
corrections for $T_{2}$ at every frequency rather than just at the peak of 
the A line. We 
tested whether the rigorous approach was necessary. For the $x=0.20$ 
sample, the effect of $T_{2}$ was nearly uniform across the entire lineshape at 150K, 
but at 30K (shown in Fig. \ref{T2_correct}(a)), 
$1/T_{2}$ increased towards the lower frequency 
side. 
Fig. \ref{T2_correct} (b) shows the integrated intensity with the
 `full $T_{2}$ correction' 
 at 150K and 30K  compared with the intensity 
points deduced from correcting for $T_{2}$ just at the peak 
frequency. Both methods gave the same results within 
experimental error, 
so we conclude that the $T_{2}$ correction at the peak of the A line 
is sufficient.

As shown in Fig. \ref{S}(a) for $x=0.12$, the intensity is constant within 
experimental error from 
300K to 70(7)K, but then shows a dramatic drop to zero from 70(7)K to 10K. 
This is the `wipeout' effect, and we will discuss the mechanism of the 
wipeout in section III. By taking 
the constant value between 300K and 70(7)K as the zero baseline and then
inverting the plot, one obtains the 
fraction of signal lost, or the wipeout fraction $F(T)$. As we 
demonstrate in section III, the onset 
temperature of $F(T)$ is in good agreement  with the charge-stripe ordering 
temperature $T_{charge}$ determined by neutron/X-ray scattering.  
Fig. \ref{Nd_wipeout} is a plot of the temperature dependence of $F(T)$ 
for all the La$_{1.6-x}$Nd$_{0.4}$Sr$_{x}$CuO$_{4}$ samples. Even 
though $T_{charge}$ and the transition widths are different for 
various $x$, all show 100$\%$ wipeout, thus leaving no observable $^{63}$Cu NQR 
signal by $\sim 10$K. We emphasize that 
since we 
have taken $T_{2}$ corrections into account for the integrated 
intensity, the loss of observable NQR 
signal is {\it not} caused by a divergence of $T_{2}$ throughout the 
entire sample as 
often observed in the vicinity of magnetic long range order. 

\subsection {$^{63}$Cu spin-lattice relaxtion}

Fig.\ref{T1} shows the spin-lattice relaxation rate
 $^{63}1/T_{1}$ for La$_{1.48}$Nd$_{0.4}$Sr$_{0.12}$CuO$_{4}$
, La$_{1.45}$Nd$_{0.4}$Sr$_{0.15}$CuO$_{4}$, and  
La$_{1.68}$Eu$_{0.2}$Sr$_{0.12}$CuO$_{4}$ in addition to
 La$_{1.885}$Sr$_{0.115}$CuO$_{4}$ data from \cite{hunt2}. 
 All $^{63}1/T_{1}$ were 
taken at A line $^{63}\nu_{Q}$, and all fit well to 
single exponential recoveries as expected for $I=3/2$ nuclei in NQR. 
The relaxation rates for the four materials 
are equal within $10 \%$ from 300K to about 125K, but below about
125K, the $^{63}1/T_{1}$ data for
 La$_{1.6-x}$Nd$_{0.4}$Sr$_{x}$CuO$_{4}$ 
measured for the remaining $^{63}$Cu NQR 
signal increases with deceasing temperature as $\sim 1/T$. The
$^{63}1/T_{1}$ data for 
La$_{1.68}$Eu$_{0.2}$Sr$_{0.12}$CuO$_{4}$ \cite{hunt2} starts 
to increase at a lower temperature of 35(5)K.
Wipeout effects below 30K 
prohibit further measurement of $^{63}1/T_{1}$ data for 
La$_{1.6-x}$Nd$_{0.4}$Sr$_{x}$CuO$_{4}$ and
La$_{1.68}$Eu$_{0.2}$Sr$_{0.12}$CuO$_{4}$.
At 30K, even though $^{63}1/T_{1}$ from the remaining $\sim 5\%$ of the signal 
is increasing with decreasing temperature, it is still measurable.

\subsection {$^{63}$Cu spin-echo decay}

The NQR spin-echo intensity $S(2 \tau)$ 
depends strongly on $2\tau$ (where $\tau$ is the time 
separation 
 between the 90$^{\circ}$ and 180$^{\circ}$ pulses) as  
 shown in 
Fig. \ref{12_T2} where $S(2 \tau)$ is plotted on a logarithmic scale.
The data is 
from the La$_{1.48}$Nd$_{0.4}$Sr$_{0.12}$CuO$_{4}$ sample, and $S(2 \tau)$  
is measured at $^{63}\nu_{Q}$ for the A line at each temperature.
In order to correct the NQR 
integrated intensity for spin echo decay, one has to extrapolate the curve 
to the zero time $2\tau =0$.
Most importantly, qualitative changes of the decay shape (such as loss 
of   
Gaussian curvature below $T_{charge}$) makes $T_{2}$ corrections 
essential in determining $T_{charge}$ accurately. 
The shortest possible $2\tau$ is limited by the  
the circuit dead time with a value of $\sim 20 \mu s$.

The fit used for the extrapolation has the functional form:
\begin{equation}
S(2\tau )=S(0)e^{-2\tau /T_{2L}}e^{-(2\tau )^{2}/2T_{2G}^{2}}
\label{T2}
\end{equation}
where $T_{2L}$ is the Lorentzian decay which has contributions from  
the Redfield $T_{1}$ process, 
and $T_{2G}$ is the Gaussian decay which is dominated by
the indirect spin-spin coupling between $like$ $spins$ \cite{slichter}. 
In various high $T_{c}$ and related copper-oxides, when the NQR linewidth 
is as small as $\sim 200$ kHz, one can excite the entire $^{63}$Cu 
NQR spectrum with intense R.F. pulses. In such cases,
 one can give 
theoretical constraints on the Lorentzian contribution $T_{2L}$ based 
on calculations of the spin-lattice relaxation process \cite{pennington}.
In the present 
case however, the full NQR linewidth ($\sim 3$ MHz) is an order of 
magnitude larger than the strength of the R.F. pulses ($\sim 200$ 
kHz). It is well known that this gives rise to artificial changes 
in the functional form of the spin-echo decay \cite{hone}, making the 
Gaussian contribution more Lorentzian. Accordingly, we chose both 
$T_{2L}$ and $T_{2G}$ as free parameters. We note that use of the 
stretched exponential form of the fitting function:
\begin{equation}
S(2\tau) = S(0)e^{-(2 \tau/T_{2})^{\beta}}
\label{T_stretch}
\end{equation}
often used in the literature under similar circumstances 
did not affect our estimate of the temperature dependence of $S(0)$. 

At $T_{charge}$ and below, the spin echo decay dramatically changed  
to a 
single rate Lorentzian. This can be seen 
by the loss of curvature in Fig. \ref{12_T2}. 
As discussed in section III, 
this crossover provides an added signature for the onset of wipeout, 
however
the crossover also generates more potential error in the estimate of 
the $T_{2}$ corrected intensity $S(0)$.
 Small changes of curvature in the fit creates large 
variations in the value of S(0) at $T\gtrsim T_{charge}$, so extra care had 
to be taken for data measurements and fits around 
$T_{charge}$.
This complexity is the reason for the larger error bars 
around $T_{charge}$ in Fig. \ref{S} (a). 
Consequently, $T_{charge}$ is given a $\pm$ 10$\%$ error.

\section{DISCUSSION}

\subsection{Wipeout effects}

We now compare the NQR wipeout fraction $F(T)$ to elastic neutron diffraction 
 studies carried out for La$_{1.48}$Nd$_{0.4}$Sr$_{0.12}$CuO$_{4}$ by  
 Tranquada {\it et al.} \cite{tranquada2}, where charge-stripe order 
 was discovered. 
 Fig. \ref{F}(a) is a comparison of the temperature dependence 
 of the square root of the transverse charge-order peak  
 \cite{tranquada2} normalized to 10K, along with our temperature 
 dependence of $F(T)$.
 We note that in general, the elastic scattering intensity 
 represents the square of the order parameter (for example, the 
 intensity of the magnetic Bragg scattering is the square of the 
 sublattice magnetization $M(T)$) \cite{lovesey}.
 The $identical$ 
 temperature dependence of $F(T)$ and the 
 neutron charge-order parameter allows us to conclude that the wipeout fraction 
 $F(T)$ represents 
 the charge-stripe order parameter. 
 Also plotted is the square root 
 of the static spin-ordered
peaks $M(T)$ for $x=0.12$ by neutron diffraction 
\cite{tranquada2,tranquada3}, and clearly indicates the wipeout fraction 
is triggered by charge-order and not spin-order. 
 
 Further evidence of charge-stripe order in 
 La$_{1.48}$Nd$_{0.4}$Sr$_{0.12}$CuO$_{4}$  \cite{zimmerman} 
and La$_{1.45}$Nd$_{0.4}$Sr$_{0.15}$CuO$_{4}$ \cite{niemoller} has been
reported using high energy x-ray scattering. 
Fig. \ref{F}(a) includes the X-ray results for $x=0.12$, where 
the square root of the longitudinal charge peaks are plotted, 
normalized to 10K. The X-ray data is in good agreement 
with both neutron and NQR data. 
Fig. \ref{F}(b)
is a comparison of X-ray data  
for $x=0.15$ \cite{niemoller}, where we have plotted the square root of the 
charge peaks in the transverse direction and compared it to our 
wipeout fraction $F(T)$ for $x=0.15$.
Again, the X-ray data agrees on the onset temperature $T_{charge}\sim 
60(6)K$ obtained from NQR. 
Both sets of X-ray data further support our identification of the 
wipeout fraction $F(T)$ as the charge-stripe order parameter. 
Furthermore, we note that $T_{charge}$ measured by NQR, X-ray, 
and neutron {\it all} agree for $x=0.12$ despite the different 
frequency scales of each probe, indicating that
charge-stripes slow down to NQR time scales very quickly near 
$T_{charge}$. 

Given the evidence for charge-stripe ordering 
in La$_{1.48}$Nd$_{0.4}$Sr$_{0.12}$CuO$_{4}$
 \cite{tranquada2,zimmerman}
 and La$_{1.45}$Nd$_{0.4}$Sr$_{0.15}$CuO$_{4}$ \cite{niemoller}, 
 we will now argue that the Cu NQR
  `wipeout' \cite{winter}
 effect 
comes as a natural consequence of the ordering. 
 Because the resonant frequency $^{63}\nu_{Q}$ is directly proportional to 
the electric field gradient (EFG) at the Cu site, we expect NQR to be 
very sensitive to local charge distributions. Indeed, a spatial 
modulation of the hole concentration ranging from 0.5 to 0 holes per 
Cu atom results in as much as an $\sim 8$ MHz \cite{yoshimura} 
instantaneous variation 
of $^{63}\nu_{Q}$. 
In the proposed stripe model \cite{tranquada1} at $x\sim 1/8$, rivers 
of hole rich CuO chains with effectively 0.5 holes per Cu 
are separated by bare three-leg CuO ladders with no holes.
We 
therefore expect that below the onset temperature for charge-stripe order
$T_{charge}$, 
regions which 
contain stripe fluctuations will have instantaneously 
varying $^{63}\nu_{Q}$ distributions, and if these fluctuations are 
on the NQR timescale (i.e. fluctuate at frequencies of order $^{63}\nu_{Q}$),
 the resonance 
condition in those regions will no longer be {\it well defined}. 
Furthermore, charge order turns on slow spin fluctuations 
\cite{tranquada4} resulting in divergences in the spin-lattice $^{63}1/T_{1}$ 
\cite{kent} and 
spin-spin $^{63}1/T_{2}$ relaxation rates within the stripe ordered 
domains (note that the measured $^{63}1/T_{1}$ and $^{63}1/T_{2L}$ 
in Fig. \ref{T1} and Fig. \ref{T2_x} 
do not necessarily reflect the 
relaxation rates in the ordered segments of the sample but they are a 
 measure of relaxation rates of the segments that have not yet 
 ordered). 
  Both these effects will result in a loss of $^{63}$Cu NQR 
signal intensity and we therefore expect the fraction of the intensity loss 
to be a good measure of stripe order. 

It is worth mentioning that at low enough temperatures,
 an NQR {\it like} signal reappears for the case of 
 La$_{1.88}$Ba$_{0.12}$CuO$_{4}$ \cite{tou}.
Below $\lesssim 10K$,
 {\it static} magnetic hyperfine fields strongly perturb the Cu NQR spectrum 
 causing a line broadening from low frequencies up to 80 MHz \cite{tou}.
 We confirmed the findings from \cite{tou} 
 for  La$_{1.88}$Ba$_{0.12}$CuO$_{4}$, where a broad, 
 featureless Zeeman perturbed NQR spectrum at 1.7K was reported. 
 We note that the low 
 temperature spectrum for La$_{1.6-x}$Nd$_{0.4}$Sr$_{x}$CuO$_{4}$ 
 is further complicated by magnetic field contributions 
 from the Nd ions that order 
 at $\lesssim 3K$ \cite{tranquada2}. 
 
We now discuss a possible fit of the temperature dependence of $F(T)$.
Conventional CDW ground states have many common characteristics with
other broken symmetry ground states such as superconducting and spin 
density wave ground states. 
In particular, within the weak coupling limit, the thermodynamics of the 
phase transition and the temperature dependence of the order parameter 
are the same as those of a BCS superconductor \cite{greuner,grunerbook}. 
To see if this is the case, we present
Fig. \ref{F}(c) where $F(t)$ is plotted as a function of the reduced 
temperature $t=T/T_{charge}$ for all the 
La$_{1.6-x}$Nd$_{0.4}$Sr$_{x}$CuO$_{4}$ samples. The solid curve is 
 the BCS form of the condensate 
density $f_{d}(t)$ in the dynamic limit \cite{exp}, and the dashed 
curves are a guide for the eye. The 
 data for the $x=0.09,0.15$, and $0.20$ samples 
 look similar, and the $x=0.07$ data stands out as having the 
 broadest transition. The $x=0.25$ data looks identical to the 
 $x=0.20$ data and has been omitted from this plot. 
 The temperature dependence of $F(t)$ for 
 $x=0.12$ 
 appears to be the sharpest and shows the best agreement with $f_{d}(t)$. 
From the wipeout mechanism we have proposed, this 
is expected since $F(T)$ 
is related to
 the fraction of the CuO$_{2}$ plane where
 condensate fluctuations exist. The fact that the transition is 
 sharpest
 around $x\sim 1/8$ could indicate that the charge-stripe is most stable around 
 $x\sim 1/8$.
 
 We now present  
 the phase diagram of La$_{1.6-x}$Nd$_{0.4}$Sr$_{x}$CuO$_{4}$ 
 in Fig. \ref{phase}, with $T_{charge}$, 
$T_{spin}$ according neutron scattering \cite{ichikawa},
$T_{spin}$ obtained by  $\mu$SR \cite{nachumi}, the 
superconducting transition temperature $T_{c}$, and the LTO to 
LTT (low temperature tetragonal) or LTLO (low temperature less 
orthorhombic) transition temperature $T_{LTT}$ \cite{crawford}.
We measured the bulk magnetic susceptibility with a SQUID 
magnetometer in a field of 10 Oe, and we 
estimated $T_{c}$ 
by taking the slope at the half point in the diamagnetic response and 
extrapolating to zero susceptibility. The $x=0.12$ and $0.07$ samples show  
residual superconducting components, and the $x=0.25$ 
sample shows no diamagnetic response down to 3K. The 
$x=0.09,0.15$, and $0.20$ samples have the largest low temperature 
susceptibility  $-4\pi \chi$ with Meissner fractions
 corresponding to $\sim 40 \%$. 
 Fig. \ref{phase} demonstrates for the first time 
that charge-stripe order exists throughout the entire
superconducting region of La$_{1.6-x}$Nd$_{0.4}$Sr$_{x}$CuO$_{4}$.
Furthermore, the $100 \%$ wipeout indicates that the charge-stripe 
transition involves the $entire$ CuO$_{2}$ plane.

Fig. \ref{phase} clearly indicates 
that $T_{charge}$ for La$_{1.6-x}$Nd$_{0.4}$Sr$_{x}$CuO$_{4}$ 
does not strictly coincide with $T_{LTT}\sim 60$K except at $x=0.12$, 
suggesting that the LTO-LTT structural transition is not 
the primary cause of the charge anomaly at $T_{charge}$. 
Further support of this statement comes from 
NQR wipeout measurements we made on 
La$_{1.8-x}$Eu$_{0.2}$Sr$_{x}$CuO$_{4}$ with $x$ = 0.07, 0.12, 0.16, and 
0.20 all with LTO-LTT structural transition temperature $T_{LTT}\sim 
130$K \cite{kataev} that are higher
than for La$_{1.6-x}$Nd$_{0.4}$Sr$_{x}$CuO$_{4}$ where $T_{LTT}\sim 60$K. 
We used exactly the same techniques to measure $F(T)$ and $T_{charge}$ 
for La$_{1.8-x}$Eu$_{0.2}$Sr$_{x}$CuO$_{4}$ as 
described earlier for La$_{1.6-x}$Nd$_{0.4}$Sr$_{x}$CuO$_{4}$, and the 
results shown in Fig. \ref{phase} indicate that 
$T_{charge}(x)$ for La$_{1.8-x}$Eu$_{0.2}$Sr$_{x}$CuO$_{4}$ 
and La$_{1.6-x}$Nd$_{0.4}$Sr$_{x}$CuO$_{4}$ 
is the same within the experimental error.
This clearly shows that $T_{charge}$ and $T_{LTT}$ are not strictly 
 correlated and that the LTT structural transition is not the primary 
 origin of the charge anomaly.
We also note that spin-stripe order has been observed in 
La$_{2-x}$Sr$_{x}$CuO$_{4}$ ($x=0.12$
\cite{suzukip,kimura} and $x=0.05$ \cite{waki}) and 
La$_{2}$CuO$_{4+\delta}$ \cite{lee},
where both materials do not even have the LTO-LTT structural phase transition.

\subsection{$^{63}$Cu spin-lattice relaxation rate $^{63}1/T_{1}$}

We now compare the temperature dependence of $^{63}1/T_{1}$ in
La$_{1.48}$Nd$_{0.4}$Sr$_{0.12}$CuO$_{4}$,
La$_{1.45}$Nd$_{0.4}$Sr$_{0.15}$CuO$_{4}$,
La$_{1.68}$Eu$_{0.2}$Sr$_{0.12}$CuO$_{4}$ 
 and La$_{1.885}$Sr$_{0.115}$CuO$_{4}$ \cite{hunt2} in light of previous 
$^{63}1/T_{1}$ measurements by Itoh 
  {\it et al.} \cite{itoh}
on a variety of the rare earth (R) doped 1-2-3 materials
  RBa$_{2}$Cu$_{3}$O$_{7-y}$ [R = Y, Nd, Eu].
 
As discussed in \cite{itoh}, the 4f 
electron moments from the trivalent rare earth R$^{3+}$ ions
contribute to the $^{63}1/T_{1}$ primarily through a dipole 
interaction between the 4f moment and the Cu nuclear moment, giving 
the general form:
\begin{equation}
^{63}1/T_{1}=(1/T_{1})_{4f} + (1/T_{1})_{Cu}
\label{T1_eq}
\end{equation} 
We find that the $^{63}1/T_{1}$ data for La$_{1.885}$Sr$_{0.115}$CuO$_{4}$,
 La$_{1.6-x}$Nd$_{0.4}$Sr$_{x}$CuO$_{4}$, and 
La$_{1.68}$Eu$_{0.2}$Sr$_{0.12}$CuO$_{4}$ has the same 
systematic features as the equivalent YBa$_{2}$Cu$_{3}$O$_{7-y}$,
 NdBa$_{2}$Cu$_{3}$O$_{7-y}$, and EuBa$_{2}$Cu$_{3}$O$_{7-y}$ data. 

(1) La$_{1.885}$Sr$_{0.115}$CuO$_{4}$ and  
YBa$_{2}$Cu$_{3}$O$_{7-y}$ have no $(1/T_{1})_{4f}$ component,
 and $(1/T_{1})_{Cu} $ for the observable segments of the NQR signal 
 in La$_{1.885}$Sr$_{0.115}$CuO$_{4}$
 decreases with decreasing temperature \cite{imai}.
 
(2) La$_{1.48}$Nd$_{0.4}$Sr$_{0.12}$CuO$_{4}$,
 La$_{1.45}$Nd$_{0.4}$Sr$_{0.15}$CuO$_{4}$,
 and NdBa$_{2}$Cu$_{3}$O$_{7-y}$ all have Nd$^{3+}$ ions which have
  $J=9/2$ ground 
states, and a large $(1/T_{1})_{4f}$ contribution is 
apparent with decreasing temperature in both cases. The temperature dependence 
of $(1/T_{1})_{4f}$ depends on the details of the crystal field 
\cite{itoh}. $(1/T_{1})_{4f}$ for NdBa$_{2}$Cu$_{3}$O$_{7-y}$
 is temperature independent down 
to at least 1K, which is the expected behaviour in the limit 
$T_{N}\ll T\ll \bigtriangleup_{1}$ where $T_{N}=0.52K$ is the 
ordering temperature and $\bigtriangleup_{1}=140K$ is the crystal 
field splitting between the ground state and the first excited state. 
The La$_{1.48}$Nd$_{0.4}$Sr$_{0.12}$CuO$_{4}$ and
La$_{1.45}$Nd$_{0.4}$Sr$_{0.15}$CuO$_{4}$
 data shows that $(1/T_{1})_{4f}$ increases with decreasing 
temperature indicating that $\bigtriangleup_{1} <140K$ in these 
materials.

(3) La$_{1.68}$Eu$_{0.2}$Sr$_{0.12}$CuO$_{4}$
 and EuBa$_{2}$Cu$_{3}$O$_{7-y}$ both have Eu$^{3+}$ ions which have $J=0$ ground 
states and $J=1$ first excited states split by the spin-orbit constant
$\lambda$. The EuBa$_{2}$Cu$_{3}$O$_{7-y}$
 data shows negligible 
$(1/T_{1})_{4f}$ contribution at and below 50K,
consistent with the $\lambda=450K \gg T$ 
\cite{itoh}. 
Since the first excited state is predominantely split by spin-orbit 
effects, one also expects negligible
$(1/T_{1})_{4f}$ contributions in La$_{1.68}$Eu$_{0.2}$Sr$_{0.12}$CuO$_{4}$
 below 50K. However, as the
La$_{1.68}$Eu$_{0.2}$Sr$_{0.12}$CuO$_{4}$
data indicates, there is an increase in $^{63}1/T_{1}$ below 35(5)K, 
suggesting that the observable $^{63}$Cu NQR signal senses diverging 
low frequency spin fluctuations due to critical slowing down of Cu 
moments toward $T_{spin}=25\sim 27K$ as observed by $\mu$SR 
\cite{wagener1} for La$_{1.68}$Eu$_{0.2}$Sr$_{0.12}$CuO$_{4}$. 
Both $\mu$SR and NQR have 
similar inherent probing frequencies of $\sim$$10^{7}$ Hz, thus one 
expects $T_{spin}$ even for a {\it glassy} transition to be comparable for 
both measurements, as is the case
 for La$_{2-x}$Sr$_{x}$CuO$_{4}$ according to $\mu$SR 
\cite{kumagi,niedermayer} and $^{139}$La 
NQR \cite{chou,suzukip,hunt2}.  
We also find that $^{63}1/T_{1}T$ in 
La$_{1.68}$Eu$_{0.2}$Sr$_{0.12}$CuO$_{4}$ decreases slightly below 
$T_{charge}$ prior to the onset of critical divergence of 
$^{63}1/T_{1}T$ toward $T_{spin}$.  Even though reduction of 
$^{63}1/T_{1}T$ is usually attributed to pseudogap in the spin 
excitation spectrum, it is not clear whether that is the case here.  
We emphasize that the reduction of $^{63}1/T_{1}T$ is found for the 
observable fraction of the CuO$_{2}$ planes, and that
the unobservable fraction may have divergent $^{63}1/T_{1}$ at 
$T \lesssim T_{charge}$.

We would like to add that although in the case of 
 La$_{1.6-x}$Nd$_{0.4}$Sr$_{x}$CuO$_{4}$ below $T=125K$,
 $(1/T_{1})_{4f}$ dominates over $(1/T_{1})_{Cu}$ and no 
 spin-ordering can be inferred, $\mu$SR measurements report that 
there is spin-ordering at temperatures $T_{spin}$ similar to   
La$_{1.8-x}$Eu$_{0.2}$Sr$_{x}$CuO$_{4}$.
$\mu$SR measurements were obtained for 
La$_{1.48}$Nd$_{0.4}$Sr$_{0.12}$CuO$_{4}$ 
\cite{nachumi,luke} where $T_{spin}\sim 30K$, and a  
reduced value for 
 La$_{1.45}$Nd$_{0.4}$Sr$_{0.15}$CuO$_{4}$ of $T_{spin}\sim 25K$ 
 has also been reported \cite{nachumi,wagener2}. 
It was also shown by $\mu$SR \cite{wagener2} that $T_{spin}\sim 25K$ for 
 La$_{1.85-y}$Nd$_{y}$Sr$_{0.15}$CuO$_{4}$ is independent of the Nd 
 concentration $y$ for $0.3\leq y \leq 0.6$. 
 
 Fig. \ref{phase} includes the values of 
 $T_{spin}$ for La$_{1.6-x}$Nd$_{0.4}$Sr$_{x}$CuO$_{4}$ according to 
 neutron scattering \cite{ichikawa} and $\mu$SR \cite{nachumi}.
  No spontaneous magnetization is detected for $x=0.20$ by $\mu$SR 
  down to 4K in contrast to the static spin-stripe 
   observations by neutron scattering \cite{ichikawa,tranquada3}, where 
   $T_{spin} \sim 20$K for $x=0.20$.
    Indeed, for $x=0.10, 0.12, 0.15,$ and 
   $0.20$, $T_{spin}$ is consistently $\sim 20$K higher for neutron 
   scattering \cite{ichikawa,tranquada3} than for $\mu$SR measurements
   \cite{nachumi}. Further evidence of the glassy nature of the 
   spin-ordering comes from ESR (electron spin resonance) studies 
   by Kataev {\it et al.} \cite{kataev} for
   La$_{1.8-x}$Eu$_{0.2}$Sr$_{x}$CuO$_{4}$ at $x\sim 1/8$, where 
   they clearly observe that the Cu spin fluctuation frequency $\omega_{sf}$
    shows a strong temperature dependence below $T_{charge}$.
  
 \subsection{ $^{63}$Cu spin-echo decay $^{63}1/T_{2}$}

We now discuss the temperature dependence of the spin echo decay both 
above and below $T_{charge}$.
 Above $T_{charge}$, the spin echo decay from all the samples
were fit using the two free parameters $T_{2G}$ 
and $T_{2L}$ in Eq. (\ref{T2}).
We found that greatly reducing the excitation 
range from $\sim 200$ kHz to $\sim 50$ kHz,
 caused a significant reduction of the Gaussian 
component. This can be understood \cite{hone} in the context of 
inhomogeneous broadening, where reducing the frequency range has the 
effect of reducing the number of {\it like} spins responsible for the 
indirect spin-spin
(Gaussian) decay. The reverse situation where the frequency range is 
kept fixed but the inhomogeneous line is broadened will also reduce the 
number of {\it like} spins. Either way, detuning the spin-spin mechanism 
 \cite{hone} results in an apparent reduction of the Gaussian 
 curvature in the spin echo envelope 
 and results in
 the Gaussian contribution to the second moment appear
 more Lorentzian in character. 
 However, even if we use weak R.F. pulses and thereby eliminate the 
 Gaussian curvature in the spin-echo decay, we found that the extrapolated value 
 S(0) does not change. 
 
 Even though the NQR HWHM does not increase significantly around 
 $T_{charge}$ (Fig. \ref{12_temp}(a)) and we maintain the same 
 strength of R.F. pulses, we 
 still observed a dramatic reduction of the Gaussian 
 component at and below $T_{charge}$, indicating that there {\it is} a 
 mechanism below $T_{charge}$ that inherently eliminates the Gaussian 
 process. 
 The same mechanism that causes the wipeout can also be used to 
 explain the crossover, namely  
 that the spatial charge modulations below $T_{charge}$
 detune
 the indirect interaction by creating site to site $^{63}\nu_{Q}$ 
 variations, thereby reducing the number of {\it like} spins 
and thus eliminating the Gaussian decay.
Below $T_{charge}$, the spin-echo decay was always Lorentzian 
with $T_{2L} \ll T_{1}$ . We observed that reducing the 
spectral excitation range
in this case did {\it not} change $T_{2L}$.
  Fig. \ref{T2_x} shows the temperature dependence of $^{63}1/T_{2L}$ 
for a selection of the La$_{1.6-x}$Nd$_{0.4}$Sr$_{x}$CuO$_{4}$ samples
and La$_{1.68}$Eu$_{0.2}$Sr$_{0.12}$CuO$_{4}$ \cite{hunt}.
The data at each doping starts at $T_{charge}$ and finishes when the 
wipeout is nearly complete.  
The doping dependence of $^{63}1/T_{2L}$ shows a 
systematic decease with increasing hole doping $x$. If $^{63}1/T_{2L}$ was 
entirely
dominated by 4f moment fluctuations from the Nd$^{3+}$
ion, $^{63}1/T_{2L}$ 
would {\it not} show this doping dependence. Moreover, as argued for the 
$^{63}1/T_{1}$ data, 
La$_{1.68}$Eu$_{0.2}$Sr$_{0.12}$CuO$_{4}$ has a negligible 4f moment 
at 50K, yet within scattering, $^{63}1/T_{2L}$ at 50K is the same  
as for La$_{1.48}$Nd$_{0.4}$Sr$_{0.12}$CuO$_{4}$. 

The fact that $T_{2L} \ll T_{1}$, coupled with the fact that the Nd$^{3+}$ 
ion is not the primary source of relaxation, leads us to believe 
that within the observable domains there exist
certain magnetic hyperfine fields $H_{hf}$ at the Cu site that 
fluctuate with a correlation time $\tau_{c}$ satisfying the motional 
narrowing limit $\gamma_{n}H_{hf}\tau_{c}\ll 
1$ \cite{slichter} (where $\gamma_{n}$ is the Cu gyromagnetic 
ratio).
We infer that these fluctuations are primarily magnetic rather than 
quadrupolar by verifying that:
\begin{equation}
\frac {^{65}1/T_{2L}} {^{63}1/T_{2L}}   \sim 
\frac {^{65}\gamma^{2}_{n}}  {^{63}\gamma^{2}_{n}}=1.15
\label{ratio}
\end{equation}
The frequency dependence of $1/T_{2L}$ for the $x=0.20$ sample 
was shown in Fig. \ref{T2_correct}(a) at $T=30K\leq T_{charge}$, and
indeed, the $15 \% $ rise at lower frequency is consistent with 
primarily magnetic relaxation.
It is also known that in the motional narrowing limit \cite{slichter}, 
\begin{equation}
1/T_{2L}\sim \gamma^{2}_{n}H_{hf}^{2}\tau_{c}
\label{limit}
\end{equation}
but since $\gamma_{n}H_{hf}$ is not known, one cannot estimate 
$\tau_{c}$. 
We note however that a sliding motion of CDW's in conventional CDW conductors 
causes motional narrowing \cite{ross}.
Our present case is unconventional 
in the sense that we are observing the fraction of the material 
that has not yet ordered, and that charge order turns on slow spin 
dynamics \cite{tranquada4}.

\subsection{The temperature dependence of $^{63}\nu_{Q}$}

Fig. \ref{12_nuQ} shows the temperature 
dependence of A line $^{63}\nu _{Q}$   
for $x=0.07$ and 0.12. The closed symbols indicate temperatures 
above $T_{charge}$, and open symbols below $T_{charge}$. Also shown 
is the vertical line marking the structural transition temperature $T_{LTT}$
 from the LTO to LTT (or LTLO). 
We define $T_{\nu_{Q}}$ as the
temperature below which the rise in the A line $^{63}\nu _{Q}$ for the 
observable part of the signal starts to increase.
$T_{\nu_{Q}}=70(7)$K for $x=0.12$ , and $T_{\nu_{Q}}\approx 130(13)$K
 for $x=0.07$.
$T_{HTT}$ is also defined as the 
HTT (high temperature tetragonal) to LTO transition temperature \cite{ginsberg}. 

The measurement of the 
temperature dependence for the A line $^{63}\nu _{Q}$ for $x>0.12$ has an added 
complexity to it. As shown in Fig. \ref{1}(a), the B line 
increases in amplitude with Sr doping, and unfortunately, the 
B line $^{65}$Cu  
frequency coincides with the main  A line $^{63}$Cu. This is evident in 
Fig. \ref{T2_correct}(a) for $x=0.20$ where the shape of the main peak is 
largely due to the $^{65}$Cu B line. Accurate determination of the 
 A line $^{63}\nu _{Q}$ temperature dependence therefore implies taking 
 very careful 
 lineshapes of a decreasing signal intensity below 
 $T_{charge}$. We have however been able to determine that the A line $^{63}\nu 
 _{Q}$ for all the samples increases with decreasing temperature down 
 to $\sim10$K and 
 that the rise below $T_{\nu_{Q}}=70(7)$K is sharp for $x=0.12$.
 $x=0.09$ and $0.12$ have qualitatively the same temperature 
 dependence of the A line $^{63}\nu _{Q}$. 

There are at least two possible explanations for the sharp 
rise in $^{63}\nu _{Q}$ 
below $T_{\nu_{Q}}$ for $x=0.12$.
One is based on electronic effects. 
 If we assume that $T_{\nu_{Q}}$ is 
in the vicinity of $T_{charge}$ which is certainly true for $x=0.12$,
 one can postulate that
 regions with lower hole concentrations wipe out at higher 
temperatures in such a way that $^{63}\nu_{Q}$ will appear to increase with 
decreasing temperature (recall that regions with higher hole concentrations have 
higher $^{63}\nu_{Q}$ values).

Another possible explanation for the large
 rise in $^{63}\nu_{Q}$ arises from
structural distortion.
In order to obtain a semi-quantitative idea of how structural effects 
change $^{63}\nu_{Q}$, we calculate the electic field gradient (EFG) at the Cu 
site by a point charge lattice summation method
similar to that reported in \cite{schimizu}. The two inputs necessary 
for the EFG calculation are the positions of the ions ({\bf r$_{i}$}) 
relative to the Cu nucleus
 and the point charge of the ions (e$_{i}$). The lattice 
 components of the EFG are 
 calculated from summations such as:
 \begin{equation}
eq_{latt}^{z}= \sum_{{\bf i}} e_{i}\frac 
{(3z_{i}^{2}-r_{i}^{2})}{r_{i}^{5}}
\end{equation} 
where we included all the ions within a sphere of radius 
100\AA\ from the origin. 
The ionic charges for the A line are assigned as follows:
\begin{equation}
La(Nd):+3
\hspace{0.2in}
O_{p} :-(2-\frac {x}{2})
\hspace{0.2in}
O_{a} :-2
\hspace{0.2in}
Cu    :+2
\end{equation}
where $O_{p}$ is the planar oxygen, $O_{a}$ the apical oxygen.
The room temperature values of the lattice constants are taken from 
high resolution X-ray diffraction
\cite{moodenbaugh}. With decreasing temperature, the percentile change
 of the lattice constants are taken from \cite{suzuki}, where the 
 orthombic splitting is reported to go as:
 \begin{equation}
 [b-a](x,T) \sim [b-a]_{o}(x)\times (1-T/T_{HTT})^{2\beta}
 \end{equation} 
where $\beta = \frac {1} {3}$. The oxygen octahedron tilting angle is taken 
to have the form \cite{buchner}:
\begin{equation}
 \Phi(x,T) \sim \Phi_{o}(x)\times (1-T/T_{HTT})^{\beta}
 \end{equation} 
also with $\beta = \frac {1} {3}$. Below $T_{LTT}$, [b-a] is set to 
zero, and $\Phi$ is kept at its maximum angle $\Phi_{o}(x)$
 \cite{buchner} whose value is given below.

For $^{63}\nu_{Q}$ in units of MHz and $eq_{latt}$ in units of $emu \times 
10^{14}$,
$^{63}\nu_{Q}$ has the general empirical form  \cite{schimizu}:

\begin{equation} 
^{63}\nu_{Q}= A(x)-B\times eq_{latt}(x,T)
\end{equation}
where $A(x)$ ($\approx 70$ MHz) is the contribution from the hole in 
the 3d$_{x^{2}-y^{2}}$ orbital and 
its value was chosen to match the room temperature value of the A line
$^{63}\nu_{Q}$.
$B$ is determined by antishielding 
effects \cite{slichter} and the value is estimated to be 14.1 MHz
 from an empirical fit performed by Shimizu {\it et al.}
 on a large pool of data for different superconducting materials.   
Our calculations typically give $eq_{latt}\approx 2.7$, consistent 
with lattice EFG's for high $T_{c}$ cuprates \cite{schimizu}. 
We note that the only parameter we have adjusted to best fit the data is the 
constant $A(x)$. All the other parameters have been taken from 
other experimental sources.

We can check that the calculation is in semi-quantitative 
agreement with the data by confirming the 
monotonic rise in $^{63}\nu_{Q}$ with deceasing temperature in the 
 LTO phase (without tilting the 
octahedra, the calculation predicts that $^{63}\nu_{Q}$ {\it 
decreases} due 
to lattice shrinking). Indeed the calculation does show 
semi-quantitative agreement with the data above $T_{\nu_{Q}}$. 

An important result from the calculation is that the change from LTO to LTT 
 has little effect on $^{63}\nu_{Q}$, i.e. 
$^{63}\nu_{Q}$ is insensitive to azimuthal rotations 
of the octahedra about the c-axis. This justifies the fact that we 
have neglected any intermediate LTLO phases \cite{crawford}.
However, $^{63}\nu_{Q}$ is 
sensitive to the angle that the octahedra are tilted {\it from} the c-axis. 
Thus in order to try to account for the dramatic increase of 
$^{63}\nu_{Q}$ below $T_{\nu_{Q}}$ for the observable NQR signal
just from structural effects, we 
$further$ tilted the octahedra from the c-axis 
starting at $T_{\nu_{Q}}$ and we took the same 
temperature dependence of the angle in the LTO phase:
\begin{equation}
\Phi(x,T<T_{\nu_{Q}}) \sim  \Phi_{o}(x) + 
\Phi_{t}(x)\times (1-T/T_{\nu_{Q}})^{\beta}
 \end{equation}

The only additional 
parameters now 
are the maximum tilt angle of the extra tilt called $\Phi_{t}(x)$, 
and $T_{\nu_{Q}}$. It is clear for $x=0.12$ that 
$T_{\nu_{Q}} \sim T_{charge}=70(7)$K, and for $x=0.07$ we have also used
$T_{\nu_{Q}} \sim T_{charge}=130(13)$K. 
The lower line in each plot is with 
$\Phi_{t}(x)=0$ and the upper line is for finite $\Phi_{t}(x)$.
The rise in $^{63}\nu_{Q}$ can now be reproduced, and the angles used  
 were as follows:

We point out that the values of $\Phi_{t}(x)$ that best reproduce the 
data should only be taken as estimates. The calculation so far 
described is naturally very sensitive to the exact ion positions and 
to the 
constant $B$. For instance, using a larger $B$ will 
make $^{63}\nu_{Q}$ more 
sensitive to changes in $eq_{latt}$, and one would then use lower 
values of $\Phi_{t}(x)$ to reproduce the data.

There is no conclusive evidence either way. 
We are reminded however that neither possibility has to
involve all segments, just those that have not yet been wiped out.
Even though there have been no reports from bulk   
scattering experiments
\cite{buchner,crawford}
that there is 
further octahedron tilting at temperatures 
below $T_{LTT}$, 
 bulk scattering experiments \cite{buchner,crawford} measure the
 spatial average of the tilting angle
while $^{63}\nu_{Q}$ is a local probe of the observable segments. 

Using the same method as described above, we also calculated
 the effects charge-stripe formation alone would have
on the EFG for $x=1/8$. 
Firstly, we confirmed that adding 0.5 holes
uniformly on the planar oxygens can account for the rise in the 
A line $^{63}\nu_{Q}$ of $\sim 8$ MHz, 
in agreement 
with experimental observations in La$_{2-x}$Sr$_{x}$CuO$_{4}$ \cite{yoshimura}. 
 Next, we took the 
 conventional stripe picture \cite{tranquada1} where the holes lie 
 uniformly in the river of holes across one Cu chain separated by bare three leg ladders,
  and we predicted an NQR 
 line splitting into the three peaks corresponding to the three 
 distinct Cu sites (actually each peak was further split $\sim 
 300$ kHz into two due to perpendicular stripes from neighbouring planes).
 In the calculation, we interpreted one hole on a Cu site to mean one 
 hole distributed evenly around its four surrounding planar oxygens. 
 The highest frequency peak corresponding to the Cu site on the river 
 shifted $\sim 6$ MHz 
 above the uniformily doped case, while the lowest peak corresponding 
 to the Cu at the centre of the 3-leg ladder shifted $\sim 3$ 
 MHz below the uniform case. We also tried various charge-stripe
 configurations \cite{white}, and we found that any smoothing of the 
 charge distribution reduced the peak splitting.
 
 Contrary to these calculated results of $^{63}\nu_{Q}$ in stripes, 
{\it no} NQR line splitting for the
observable part of the signal was observed experimentally 
 at or below $T_{charge}$. This implies that there is
 no static stripe order in the {\it observable part} of the CuO$_{2}$ plane. 
 The signal we can observe below $T_{charge}$ is either from islands 
 that have not yet ordered, or from islands with quasi-static order 
but with NQR lines that are motionally narrowed. 
 
\section{conclusions} 

In this paper, we have reported a systematic study of the temperature
 and doping dependence of stripe instabilities in 
La$_{1.6-x}$Nd$_{0.4}$Sr$_{x}$CuO$_{4}$ throughout the superconducting 
regime based on $^{63}$Cu NQR. 
Our novel approach takes advantage of the extreme sensitivity of 
$^{63}$Cu NQR to charge-stripes. We have confirmed that the NQR 
wipeout fraction $F(T)$
is a good measure 
of the charge-stripe order parameter 
\cite {tranquada2,niemoller,zimmerman}, 
and we have
extended earlier measurements of the charge-stripe order parameter
based on diffraction techniques beyond $x=0.12,0.15$ 
to $0.07\leq x \leq 0.25$ and in doing so, 
obtain an extended phase diagram of the
La$_{1.6-x}$Nd$_{0.4}$Sr$_{x}$CuO$_{4}$ system. 
We have also presented and discussed the temperature and doping 
dependence of the NQR parameters 
$^{63}\nu_{Q}$, $^{63}1/T_{1}$, and $^{63}1/T_{2}$.  

We have shown that robust charge-stripe order continues to hold up to 
$x=0.25$, where no static hyperfine fields have been reported by $\mu$SR. This 
implies that completely 
static spin-ordering is {\it not} a necessity for charge-ordering.
On the other hand, the Lorentzian spin-spin relaxation rate 
$^{63}1/T_{2L}$ observed below $T_{charge}$ suggests that 
charge-stripes continue to fluctuate slowly even below $T_{charge}$. 

Our observation that $T_{charge}$ is higher for lower doping $x$ is 
counterintuitive, given that the charge-stripe transition is sharpest 
for $x=\frac{1}{8}$.  On the other hand, the lower hole concentration 
$x$, the stronger the tendency towards charge localization.  This 
might explain why the onset of charge order is as much as a factor of 
two higher in temperature for $x$=0.07 than for $x$=0.12.  We should 
also recall that NQR is a local probe sensitive to stripes no matter 
how they are disordered.

Comparison of the stripe-superconductivity phase diagram 
of La$_{1.6-x}$Nd$_{0.4}$Sr$_{x}$CuO$_{4}$ 
with our results obtained for La$_{2-x}$Sr$_{x}$CuO$_{4}$ 
\cite{hunt}, 
reveals a possibly striking feature: it would appear that charge-stripe order 
 stops when $T_{charge}$ becomes lower than $T_{c}$, which for 
 La$_{2-x}$Sr$_{x}$CuO$_{4}$ happens to occur at $x\sim 1/8$, but for
 La$_{1.6-x}$Nd$_{0.4}$Sr$_{x}$CuO$_{4}$ does not happen due to the 
 highly suppressed $T_{c}$ and the larger $T_{charge}$.
A natural question to ask is whether the 
 stripe instabilities and superconductivity compete. If one looks at 
 the La$_{2-x}$Sr$_{x}$CuO$_{4}$ with $x>1/8$, where static-stripe ordering 
 is supressed and superconductivity is robust, one may conclude they 
 compete with each other. However, inelastic neutron scattering 
 measurements \cite{tranquada4} indicate that low energy ($\gtrsim 2meV$) 
 dynamic stripe fluctuations extend beyond $x=1/8$, and perhaps they coexist 
 even in YBa$_{2}$Cu$_{3}$O$_{6.6}$ \cite{dai}. 
 Furthermore, neutron scattering on La$_{2}$CuO$_{4+\delta}$
 by Y.S. Lee {\it et al.} \cite{lee} report that the elastic spin-order 
 intensity appears at the same temperature as the superconductivity, 
 suggesting that the two phenomena are strongly correlated. 
 Our new 
 stripe-superconductivity phase diagram for 
 La$_{1.6-x}$Nd$_{0.4}$Sr$_{x}$CuO$_{4}$ in Fig. \ref{phase} also 
 clearly indicates that the superconductivity phase boundary exists within 
 the charge stripe phase boundary.
 
 \subsection{ACKNOWLEDGMENTS} 
 
 We thank K.R. Thurber, Y.S. Lee, R.J. Birgeneau, M.A. Kastner, J. 
 Tranquada, and V.J. Emery for valuable discussions. This work was 
 supported by NSF DMR 99-71266, NSF DMR 98-08941, and in part by the 
 A.P.Sloan foundation and the Mitsui foundation. 
 One of us (P.M.S.) has been supported by the Platzmann fund during the
course of this study.

\begin{figure} 
\caption{(a) Lineshapes for La$_{1.6-x}$Nd$_{0.4}$Sr$_{x}$CuO$_{4}$ where 
the value 
$0.07\leq x \leq 0.25$ is shown for each line.
 All lineshapes were taken at 150K and all are 
normalised to equal heights for purposes of comparison.
 The raw data points are shown along with 
their fits (see text) and the decomposition of the fit for $x=0.12$ is shown.
(b) $^{63}$Cu A line HWHM ($\blacktriangle$) obtained 
by fitting the 150K lineshapes.
(c) A line $^{63}\nu_{Q}$ ($\bullet$), B line 
$^{63}\nu_{Q}$ ($\circ$) both at 150K.}
\label{1}
\end{figure}

\begin{figure}
\caption {(a) Temperature 
dependence of the $^{63}$Cu A line HWHM for $x=0.12$ ($\circ$) along 
with a linear 
extrapolation, and the temperature dependence of
A line $^{63}\nu_{Q}$ for $x=0.12$ ($\bullet$). 
(b) $x=0.12$ lineshapes at 
150K ($\circ$), 77K ($\bullet$), 60K ($\triangle$), 40K ($\blacktriangle$), all 
corrected for Boltzmann factor (see text). }
\label{12_temp}
\end{figure}

\begin{figure}
\caption {(a) Temperature dependence of the integrated 
$^{63}$Cu NQR intensity for 
La$_{1.48}$Nd$_{0.4}$Sr$_{0.12}$CuO$_{4}$ 
in arbitrary units. Also shown, the onset temperature for wipeout 
$T_{charge}$, and the ($\circ$) show Gaussian curvature in the spin 
echo decay, ($\bullet$) do not (see text).
(b) Temperature dependence of the integrated intensity for 
La$_{1.875}$Ba$_{0.125}$CuO$_{4}$
 for the A line ($\triangle$) and B line ($\blacktriangle$).}
\label{S}
\end{figure}

\begin{figure}
\caption {(a) $^{63}1/T_{2}$ in $ms^{-1}$ ($\bullet$) as a function of 
frequency at 30K for the $x=0.20$ sample. 
Also shown, the 30K lineshape in arbitrary units measured at fixed 
$\tau=0.12\mu s$ ($\circ$) along with its fit and the decomposition of 
the fit. 
(b) Temperature dependence of the integrated intensity for 
$x=0.20$ measured by the full $T_{2}$ correction 
 ($\bullet$), or by correcting for $T_{2}$ just at the peak of 
 the A line ($\times$). }
\label{T2_correct}
\end{figure}

\begin{figure}
\caption{Wipeout fraction $F(T)$ in 
La$_{1.6-x}$Nd$_{0.4}$Sr$_{x}$CuO$_{4}$ for
$x=0.07$ ($\blacktriangle$),
$x=0.09$ ($\triangle$), $x=0.12$ ($\bullet$), 
$x=0.15$ ($\circ$),$x=0.20$ ($\blacktriangledown$), and $x=0.25$ ($\diamond$).
The solid lines through the points are a guide for the eye.}
\label{Nd_wipeout}
\end{figure}

\begin{figure}
\caption {$^{63}1/T_{1}$ temperature dependence
 for La$_{1.48}$Nd$_{0.4}$Sr$_{0.12}$CuO$_{4}$ ($\bullet$),
La$_{1.45}$Nd$_{0.4}$Sr$_{0.15}$CuO$_{4}$ ($\triangle$), 
La$_{1.68}$Eu$_{0.2}$Sr$_{0.12}$CuO$_{4}$ ($\circ$)
 and the solid line 
La$_{1.885}$Sr$_{0.115}$CuO$_{4}$ $^{22}$,
all measured in $ms^{-1}$.}
\label{T1}
\end{figure}

\begin{figure}
\caption {Spin echo decay for La$_{1.48}$Nd$_{0.4}$Sr$_{0.12}$CuO$_{4}$
at the various temperatures 
normalized by the Boltzmann factor. Also shown,
 are spin echo envelopes fit with 
Gaussian curvature ($\circ$), and without ($\bullet$).}
\label{12_T2}
\end{figure}

\begin{figure}
\caption{(a) Plots of the wipeout fraction $F(T)$ ($\circ$),
  the neutron charge order parameter $^{7}$ 
($\bullet$), the X-ray order parameter $^{8}$ ($\blacktriangle$), 
and the neutron spin order parameter $^{7}$ ($\times$) for $x=0.12$. Each 
data set is normalized to its 10K value and
the dotted lines are guides for the eye.
(b) Wipeout fraction $F(T)$ 
($\circ$) along with the X-ray order parameter $^{9}$ ($\bullet$) for $x=0.15$.
Each 
data set is normalized to its 10K value and
the dotted line is a guide for the eye.
(c) Wipeout fraction $F(t)$ as a function of the reduced 
temperature $t=T/T_{charge}$ for $x=0.07$ ($\triangledown$) 
, $x=0.09$ ($\circ$), $x=0.12$ ($\bullet$), $x=0.15$ 
($\triangle$), and $x=0.20$ ($\blacktriangle$).  
The solid line is $f_{d}(t)$, and the 
dashed lines are guides for the eye. }
\label{F}
\end{figure}

\begin{figure}
\caption{Phase diagram with $T_{charge}$ for La$_{1.6-x}$Nd$_{0.4}$Sr$_{x}$CuO$_{4}$
($\bullet$), $T_{charge}$ for 
La$_{1.8-x}$Eu$_{0.2}$Sr$_{x}$CuO$_{4}$ ($\circ$), $T_{spin}$ for La$_{1.6-x}$Nd$_{0.4}$Sr$_{x}$CuO$_{4}$ 
according to neutron scattering ($\times$) $^{12}$,
$T_{spin}$ for La$_{1.6-x}$Nd$_{0.4}$Sr$_{x}$CuO$_{4}$ according to $\mu$SR
 ($+$) $^{14}$, 
and $T_{c}$ for La$_{1.6-x}$Nd$_{0.4}$Sr$_{x}$CuO$_{4}$
($\blacktriangle$) along with a shaded region to highlight the
superconducting phase.
Also shown is the LTO to LTT (or LTLO) boundary (broken line)
 for La$_{1.6-x}$Nd$_{0.4}$Sr$_{x}$CuO$_{4}$ $^{32}$,
and two dashed lines joining the ($\times$) and ($+$) points as a guide 
for the eye.
Near the solubility limit, we find $T_{charge}=40(10)$K for La$_{1.35}$Nd$_{0.4}$Sr$_{0.25}$CuO$_{4}$
(not shown) to be about the same as $T_{charge}=40(6)$K for La$_{1.40}$Nd$_{0.4}$Sr$_{0.20}$CuO$_{4}$.}
\label{phase}
\end{figure}

\begin{figure}
\caption {Temperature dependence of $^{63}1/T_{2}$ in $ms^{-1}$
for La$_{1.6-x}$Nd$_{0.4}$Sr$_{x}$CuO$_{4}$ with
 $x=0.07$ ($\bullet$), $x=0.12$ ($\circ$),
 $x=0.25$ ($\triangle$), and also for  
La$_{1.68}$Eu$_{0.2}$Sr$_{0.12}$CuO$_{4}$ ($\blacksquare$).}
\label{T2_x}
\end{figure}

\begin{figure}
\caption {Temperature dependence of $^{63}\nu_{Q}$ A line for (a) 
$x=0.07$ and
 (b) $x=0.12$. ($\bullet$) corresponds to data 
for $T > T_{charge}$ and ($\circ$) to $T < T_{charge}$. The vertical 
line indicates the LTO-LTT transition temperature.
The upper curve of the fit corresponds to finite $\Phi_{t}(x)$, 
and the lower curve to $\Phi_{t}(x)=0$.}
\label{12_nuQ}
\end{figure}

\begin{table}
\caption{}
\begin{tabular}{ddd}
$x$ & $\Phi_{o}(x)$ & $\Phi_{t}(x)$ \\ \hline
0.07   &  5.5$^{o}$  &   4.5$^{o}$ \\
0.12   &  4.8$^{o}$  &   4.5$^{o}$ \\ 
\end{tabular}
\end{table}


\begin{references}
\bibitem{exp} We note that in the BCS dynamic limit, the 
 temperature dependence of $f_{d}(T)$ is almost identical to that of 
 the order parameter itself \cite{grunerbook}, but it would be 
 physically more accurate to associate  
$F(T)$ with $f_{d}(T)$.
\end{references}
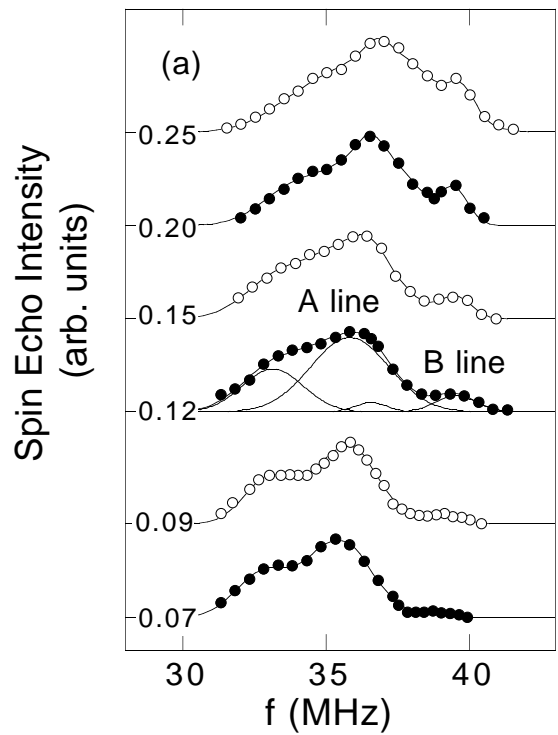
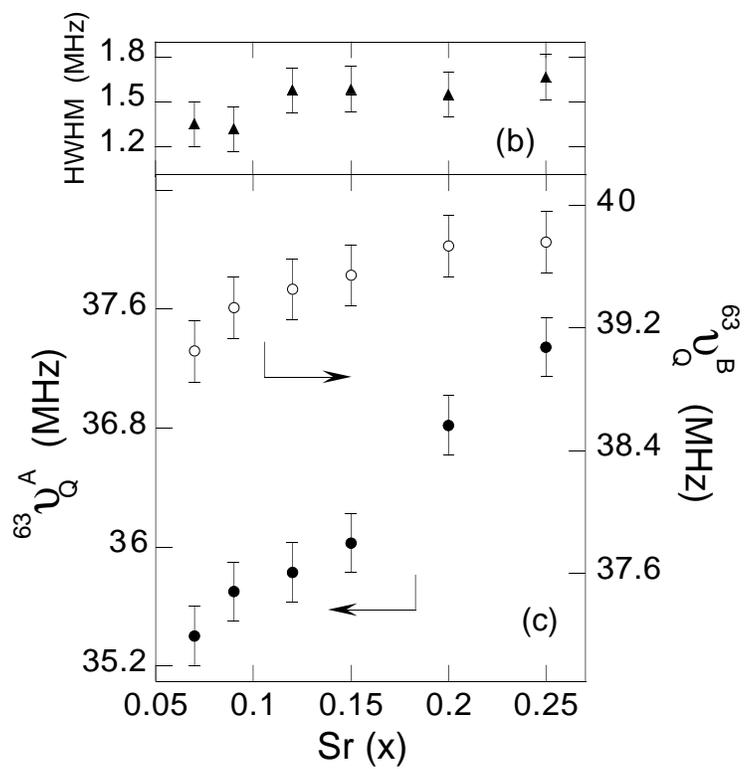
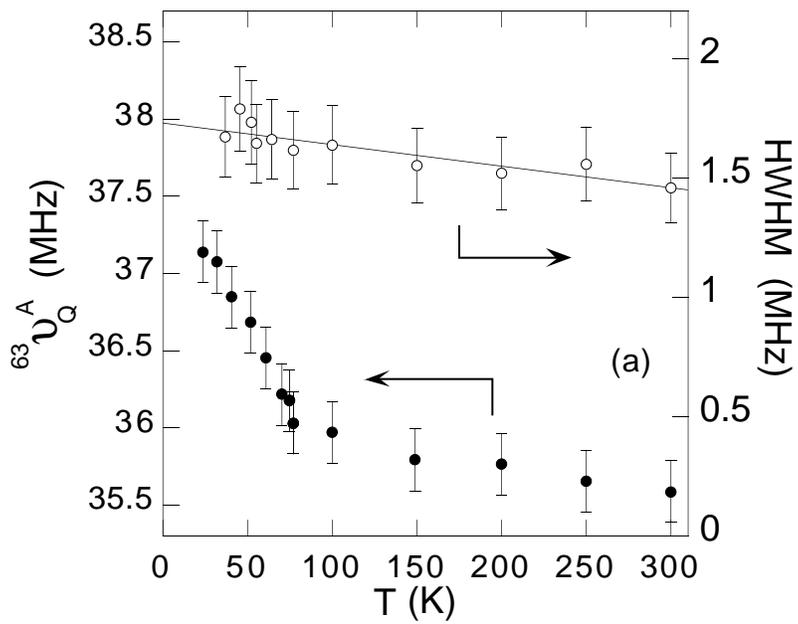
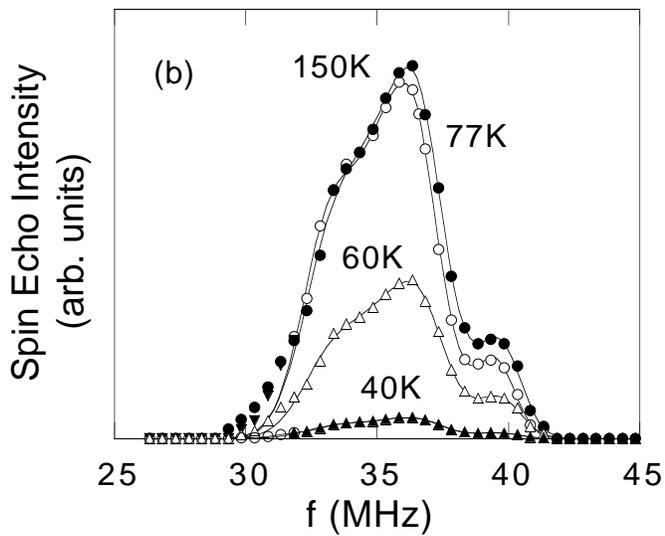
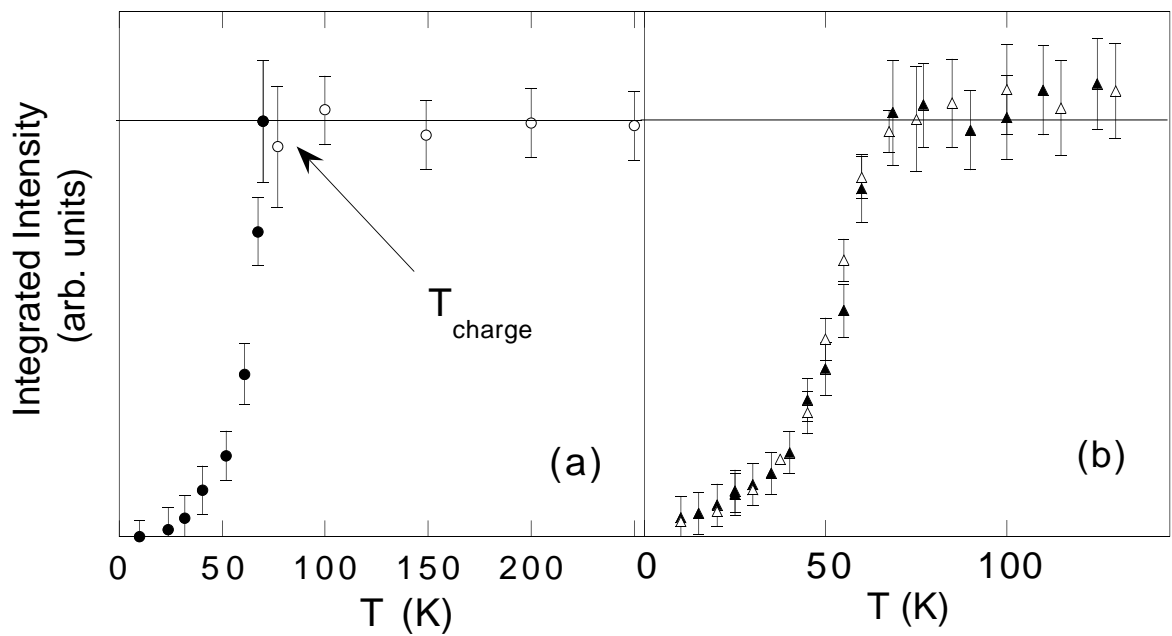
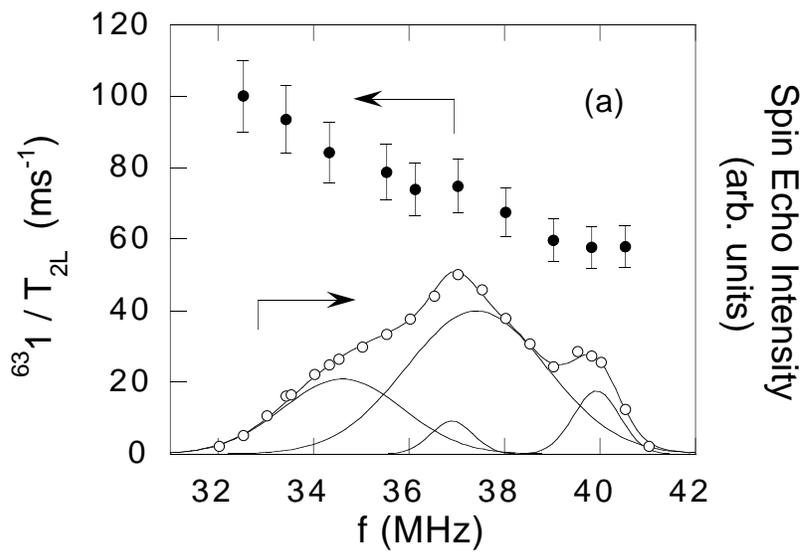
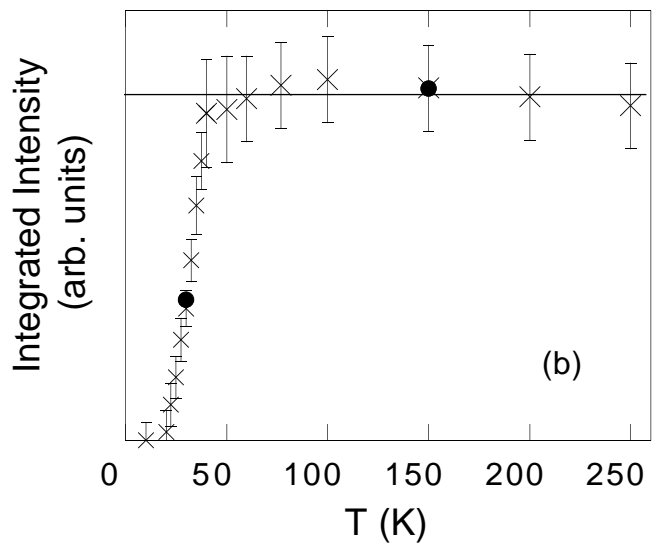
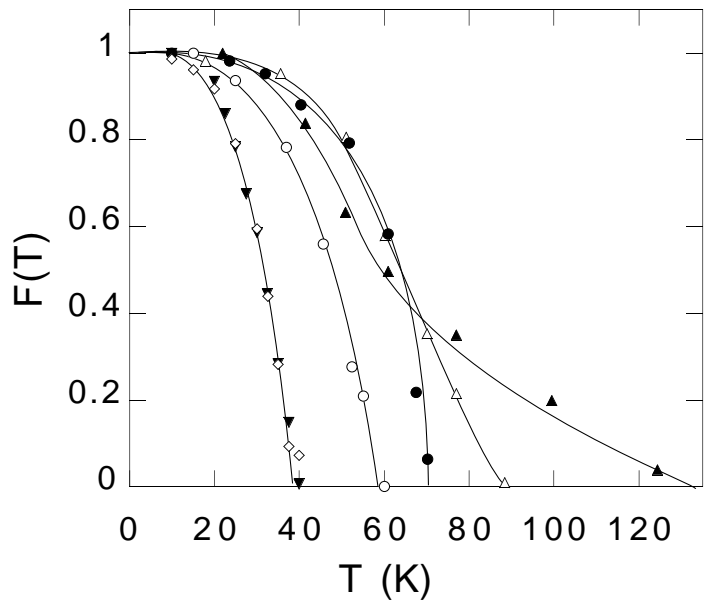
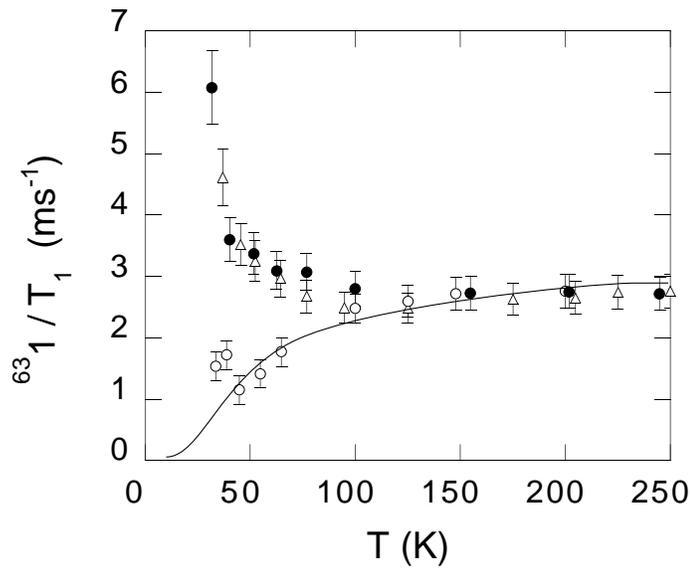
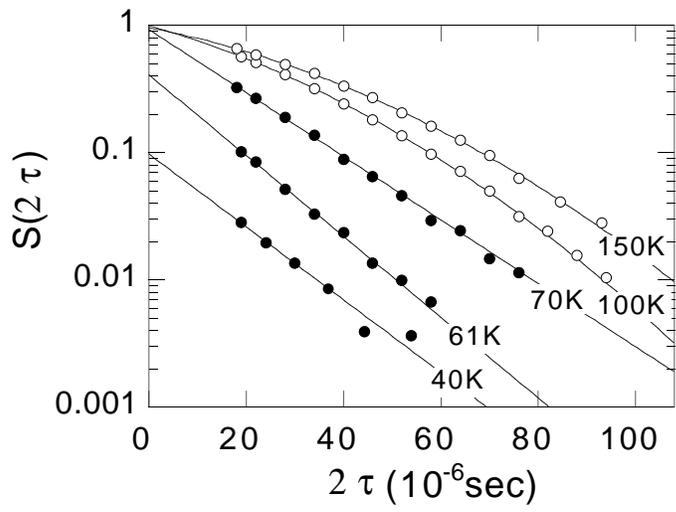
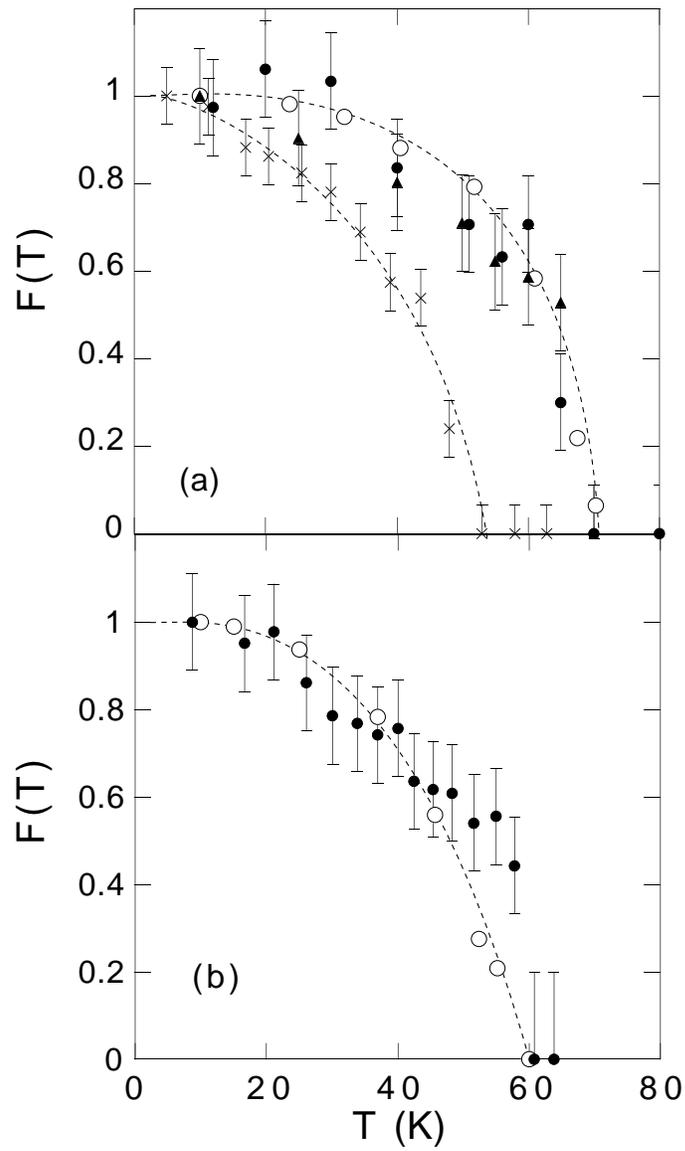
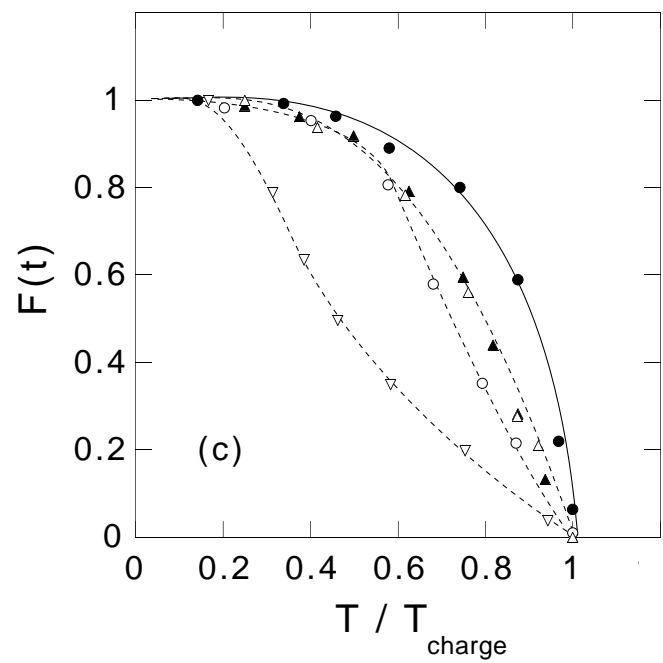
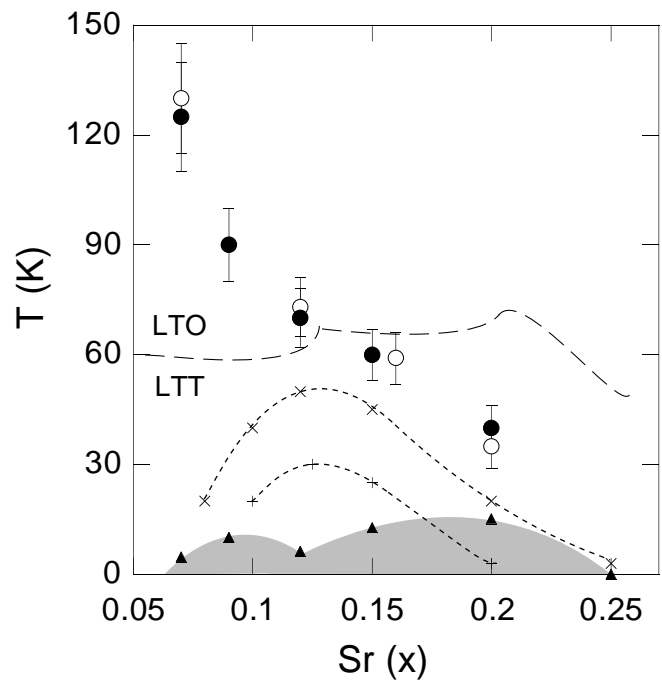
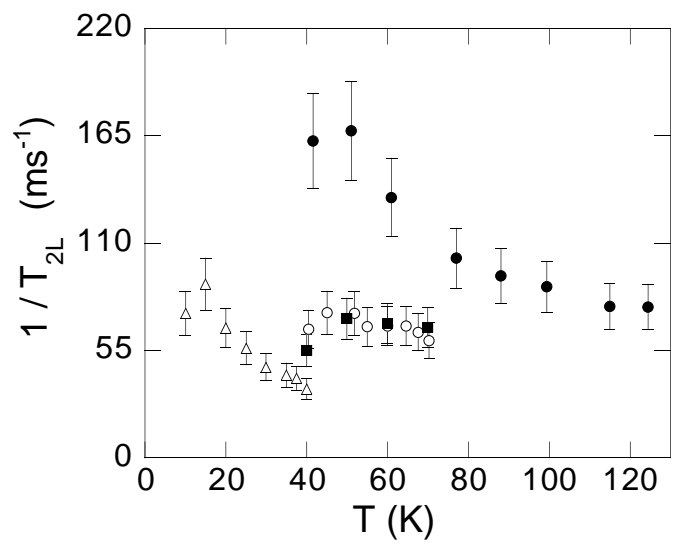
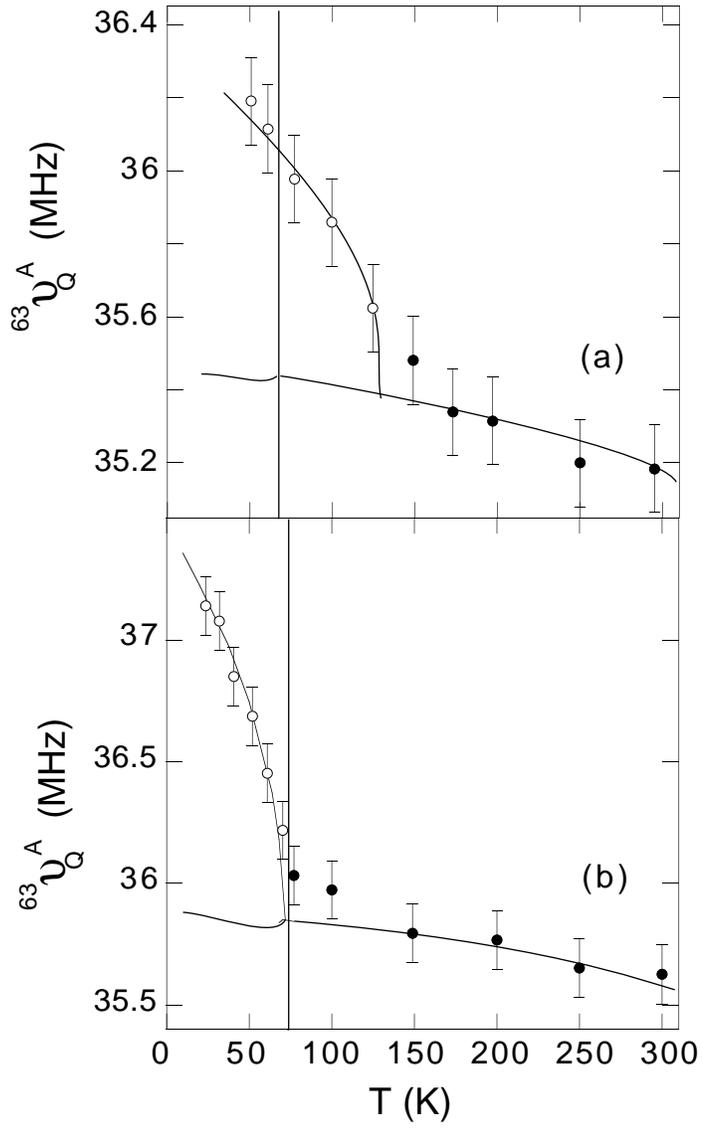
\end{document}